\documentclass[12pt]{article}
\usepackage{graphics} 
\usepackage{latexsym} 
\newcommand{\fr}{\frac} 
 
\newcommand{\ra}{\rightarrow} 
\newcommand{\beq}{\begin{eqnarray}} 
\newcommand{\enq}{\end{eqnarray}}

\newcommand{\w}{\omega}
\newcommand{\e}{\epsilon}
\newcommand{\p}{\partial}
\newcommand{\te}{\theta}

\newcommand{\vp}{\varphi}
\newcommand{\be}{\begin{equation}} 
\newcommand{\en}{\end{equation}} 
\newcommand{\ee}{\end{equation}}
\newcommand{\no}{\nonumber}

\begin{document} 
\bigskip \bigskip \bigskip 
\centerline{ \bf  \large \large On the stability of the quantum hall soliton.} 
\bigskip 

\bigskip 
\centerline{\bf Iosif Bena and Aleksey Nudelman } 
\medskip 
\centerline{Department of Physics} 
\centerline{University of California} 
\centerline{Santa Barbara, CA  93106-9530 U.S.A.} 
\medskip 
\centerline{email:iosif and anudel@physics.ucsb.edu } 
\bigskip \bigskip

\begin{abstract}
In this note we investigate the stability of the classical ground state of the Quantum Hall Soliton proposed recently in hep/th 0010105. We explore two possible perturbations  which are not spherically symmetric and we find that the potential energy decreases in both case. This implies that the system either decays or is dynamically stabilized (because of the presence of magnetic fields). 
If one makes an extra  assumption that in the real quantum treatment of the problem string ends and D0 branes move together (as electrons and vortices in the Quantum Hall effect), a static equilibrium configuration is possible.
\end{abstract}

\section{Introduction}

Recently Bernevig, Brodie, Susskind and Toumbas \cite{bbst} proposed a very interesting string theoretical description of a two dimensional electron system. The brane setup consists of a D2 brane wrapped on a 2-sphere in the near horizon limit of a large number $K$ of D6 brane. By the Hanany-Witten effect \cite{hw}, $K$ strings are extended from the D6 branes to the D2 branes. To make the configuration stable, $N$ D0 branes, which are repelled by D6 branes, are dissolved into the D2 brane. This brane setup, called the Quantum Hall Soliton, describes a 2 dimensional system of charged particles (string ends) in a large magnetic flux coming from the D0 branes. The system exhibits several interesting phenomena, similar to those found in real quantum Hall systems.

In \cite{bbst} the stability of the Quantum Hall Soliton was examined with respect to D0 brane emission, D2 brane nucleation as well as spherically symmetric perturbations of potential energy. The purpose of the paper is to examine the stability of the QH soliton under perturbations without spherical symmetry. In particular we will consider a deformation in which the D2 sphere is slightly distorted to form an ellipsoid, as well as a configuration in which the center of the D2 sphere is moved slightly from the position of the D6 branes.

Intuitively both systems should be unstable, because the string ends will tend to concentrate in the regions close to the D6 branes (since they will have less energy there), and the D0 branes will tend to concentrate in the regions far away from the D6 brane (because of D0-D6 repulsion). Thus, regions which are close to the D6 branes will be pulled even closer by the strings, and regions which are far away will be pushed even further. However, this naive argument ignores the repulsion of the strings ends and of D0 branes, as well as the effect of the D2 brane tension, which could in principle compensate the D0/D6 repulsion and F1/D6 attraction. Therefore one needs to make a careful analysis of the physical effects involved when there is a small non spherical perturbation in the system.

\section{The Physics}

In this section we will explain the physical phenomena which take place when a QH soliton is deformed.  We will keep our discussion general, and give more details in  the following sections. The spatial volume of D2 brane  is parametrized by coordinates $\theta$ and $\vp$. We only consider axially symmetric perturbations.

The distribution of D0 branes is given by the magnetic field strength $F_{\theta \vp}$. The DBI Lagrangian is a functional of  $F_{\theta\vp}$, and also depends on  the embedding of the D2 brane in spacetime.

When the shape of the D2 brane changes, the embedding changes as well, and thus the form of the functional changes. One needs to find the $F_{\theta\vp}$ which minimizes the new functional, with the constraint that the integral of  $F_{\theta\vp}$ (which gives the total number of D0 branes) is fixed.

One also needs to find the distribution of the string ends on the D2 brane. In the spherically symmetric case, there is a constant positive charge density on the D2 brane, coming from the pullback of the spacetime 2-form magnetically sourced by the D6 branes. The total induced positive charge is thus $K$, and it is neutralized by the $K$ strings. Because of the spherical symmetry, the negative charge density of the string ends cancels the induced charge density everywhere, and thus there is no electric field on the brane. 

In the axially symmetric case, the induced positive charge becomes a function of $\te$. Moreover, since the length of the strings  now depends on $\te$, there is an effective force which pulls the string ends towards where the strings are shorter. In the equilibrium state, there is an electric force which equilibrates the force coming from the string tension. One can find this force easily. If $T_{NG}(\te)$ is the Nambu-Goto energy of a string with the end at position $\te$ on the D2 brane, the electric field on the brane will be
\be
E_{\te} = {\p T_{NG}(\te) \over \p\te}.
\label{Felectric}
\ee
Naively, one can also use Gauss's law to get the total charge density
\be
\rho_{gauss} = d * E,
\ee
where $E$ is understood as a one form. This charge density is the sum of the induced charge density and the charge density of the string ends. We will see later that taking into account the large D0 charge of the D2 brane modifies quantitatively, but not qualitatively this density. The density of string ends is given by
\be
\rho_{gauss}= \rho_{strings} + \rho_{induced}.
\ee
The integral of the string tension with the string density gives the contribution  of the strings to the energy. One needs however not to forget to take into account the new electric field in the Born Infeld action.

\begin{center}
\begin{figure}
\scalebox{0.7}{\includegraphics{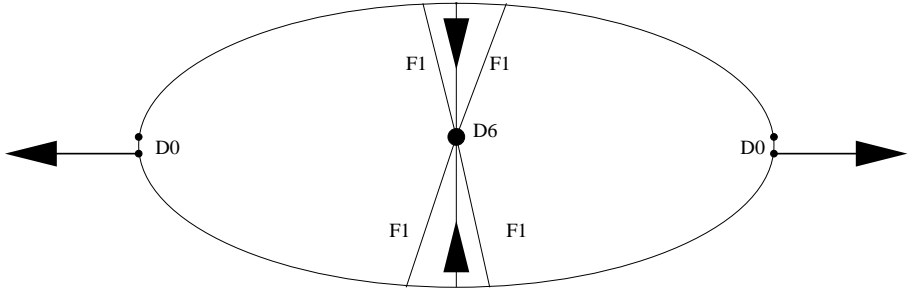}}
\caption{Naive instability of the proposed Quantum Hall system }
\end{figure}
\end{center}

\section{Embedding}
We rescale our coordinates like in \cite{bbst} in order to make the computations more transparent:
\be
y^{\mu} \ra (K g_s/2)^{1/3} y^{\mu},\ \ \ x^i \ra (K g_s/2)^{-1/3} x^i,
\ee
where $y^{\mu}$ and $x^i$ denote coordinates parallel and transverse to the D6 brane respectively. In these coordinates the near horizon D6 brane metric, dilaton and 2-form are
\beq
ds^2&=&\sqrt{\fr{\rho}{l_s}}\eta_{\mu\nu}dy^{\mu}dy^{\nu}-\sqrt{\fr{l_s}{\rho}}dx^idx^i, \nonumber \\
e^{-\phi}&=&\fr{g_s K}{2} \left ( \fr{l_s}{\rho} \right )^{3/4}, \nonumber \\
H^{(2)}&=&{K l_s \over 4 \rho^3}\epsilon_{ijk}x^i dx^j \wedge dx^k,
\label{metric}
\enq
where the $\e$ symbol is numerical, and $\rho^2 \equiv (x^1)^2+(x^2)^2+(x^3)^2$.

\subsection{Ellipsoidal deformation}

Let us consider the deformation of the spherical D2 brane into an ellipsoid. The new embedding into spacetime is
\beq
x^1=r\sin \theta \cos \vp,\ \ \  
x^2=r\sin \theta \sin \vp, \ \ \
x^3=r\sqrt{1+\e} \cos \theta, 
\enq
where $\e$ is a small parameter. The induced metric and 2-form on the D2 brane are
\beq
G_{00}&=&\sqrt{\fr{\rho}{l_s}},\ \ \ G_{\te\te}=-r^2(1+\epsilon \sin^2 \theta ) \sqrt{\fr{l_s}{\rho}}, \nonumber \\
G_{\vp\vp}&=&-\sin^2 \te r^2 \sqrt{\fr{l_s}{\rho}},\ \ \ 
H_{\theta \vp}=\frac{K l_s\sqrt{1+\epsilon} \sin \theta}{2(1+\epsilon \cos^2 \theta)^{3/2}},
\label{indmetric}
\enq
where the distance from the D6 branes is $\rho=r\sqrt{1+\epsilon \cos^2 \theta}$.

The D2 brane also carries $N$ units of D0 brane charge, 
\be
F_{\te\vp} = {N\over 2} f(\te), \ \ \ \int_0^{\pi}f(\te) d \te=2.
\label{F}
\ee
As explained before, in the spherically symmetric case $f_0(\te)= \sin \te$, but now it is a general function which needs to be determined by minimizing the DBI action of the D2 branes.

\subsection{The ingredients of the action}
The action for a single  open string coupled to an electric field is \cite{burgess}
\be
S_{f_1}=-\fr{1}{2\pi l_s^2} \int_{\Sigma} d^2 \sigma \sqrt{-det ( \p_\alpha 
X^\mu \p_\beta  X^\nu g_{\mu \nu})} -\int_{\p \Sigma} ds \fr{dX^{\mu}}{ds} A_{\mu}(X),
\label{string}
\en
where $ g_{\mu \nu}$ is the target space metric (\ref{metric}), $\alpha$ and $\beta$ are worldsheet coordinates, and $A_{\mu}$ is the vector potential on the D2 brane. 
The relevant part of the D2 brane action is
\beq
S_{D2}&=&-\fr{1}{4\pi^2 g_s l_s^3} \int d\tau d\theta d \vp e^{-\phi} \sqrt{det (G_{ab}+2\pi l_s^2 F_{ab})} \nonumber \\
&+&\fr{1}{4\pi^2 l_s^3 } \int_{D2} 2\pi l_s^2 {H}^{(2)} \wedge A^{(1)},
\label{d2}
\enq
where ${H}_2$ and  $G_{ab}$ are given in (\ref{indmetric}). The components of $F_{ab}$ are both magnetic (\ref{F}) and electric  (\ref{Felectric}).

We assume there is a large number of string ends on the D2 brane, which can be approximated by the continuous distribution $\rho^{strings}_{\te\vp}$
which integrates to $K$. The potential energy coming from the strings has a part coming from the Nambu-Goto string action and another part coming from the action of the string ends. Equation (\ref{string}) gives
\beq
V_{strings}&=& \int_{ellipsoid} \rho^{strings}_{\te\vp} \left({\rho \over 2 \pi l_s^2}+ A_0\right) \nonumber \\ 
&=&  {K r \over 2 \pi l_s^2} \int_0^{\pi}{d \te \left({{\rho} \over r}+ A(\te)\right)\rho^{strings}(\te)},
\label{vstrings}
\enq
where $\rho^{string}(\te)$ is now a scalar density which integrates to 2. We also rescaled $A_0$ and integrated over $\vp$. 

Another piece of the potential comes from the Wess-Zumino term of the D2 brane. Using the form of $H_{\te\vp}$ from (\ref{indmetric}) and rescaling $A_0$ as in the formula above, the WZ term is
\be
V_{WZ}= - {K r \over 4 \pi l_s^2}  \int_0^{\pi}{d \te H(\te) A(\te)},
\label{vwz}
\ee
where $H(\te) = \sin(\te) \sqrt{1+\e} (1+\e \cos^2 \te)^{-3/2}$ is a scalar density coming from $H_{\te\vp}$ in (\ref{indmetric}). 

We can also obtain the DBI contribution to the potential by substituting the metric, dilaton (\ref{indmetric}), magnetic (\ref{F}) and electric ($\p_{\te} A_0$) field strengths:
\be
V_{DBI} = {Kr \over 4 \pi l_s^2} \int_0^{\pi}d \te \sqrt{B(\te)+C(\te)f(\te)^2+\sin ^2 \te (\p_\te A(\te))^2 },
\label{vdbi}
\ee
where $B(\te)= \sin^2 \te (1+\e \sin^2 \te)/(1+\e \cos^2 \te)$, $C(\te)=\fr{\pi^2 N^2 l_s^2}{r^3 (1+\e \cos^2 \te)^{1/2}}$, and $A(\te)$ is defined above. To O($\e^0$) we obtain the spherically symmetric case: $B_0(\te)= \sin^2 \te $, $C_0= \pi^2 N^2 l_s^3/r^3 \equiv c$, $f_0(\te)^2 = \sin \te$, and $A(\te)$ is constant. Thus the DBI term in the action has a contribution from  $A$ only to $O(\e^2)$.
One can now minimize the total potential to find the form of $A(\te)$, $f(\te)$ and $\rho^{strings}(\te)$, with the constraints that $f(\te)$ and $\rho^{strings}(\te)$ integrate to 2.

One obtains:
\beq
A(\te)&=& - {\rho \over r} \label{main1}, \\
\rho^{strings}(\te) - H(\te) &=& \p_{\te} {\sin^2 \te \p_{\te} A(\te) \over \sqrt{B(\te)+C(\te)f(\te)^2+\sin ^2 \te (\p_\te A)^2 }} \label{main2}, \\
f(\theta)^2 &=& {B+ \sin ^2 \te (\p_\te A)^2 \over C^2/\lambda^2 - C},
\label{main3}
\enq
where  $\lambda$ is the Lagrange multiplier which enforces the constraint on $f(\te)$. Imposing the constraint determines $\lambda$. 
In Chapter 2, we argued that one can find the electric field on the brane by differentiating the $\theta$ dependent tension of a stretched string. This is equivalent to equation (\ref{main1}). We also argued that the total charge density, which is the sum of the string density and the induced density (proportional to $-H(\te)$) can be obtained by applying Gauss's law. Here we see that the argument was a bit naive since it did not take into account the fact that a significant part of the energy came from magnetic flux. Equation (\ref{main2}) remedies that.

The first observation one can make is that (\ref{main1}) implies that $V_{strings} = 0$. Thus, one does not need to compute $\rho^{strings}$ any more. Moreover, we can argue that in order to  find the $O(\e^2)$ correction to the energy one only needs to compute the first order change in $f$. This can be seen by expanding
\be
f(\theta)=\sin \te + \e f_1(\te) + \e^2 f_2(\te).
\ee
The $\e^2$ contribution to the energy coming from $f_2$ is proportional to $\int f_2(\te) d \te$, which is $0$ by the constraint (\ref{F}). Thus one only  needs $f_1(\te)$, which after a few steps is found to be
\beq
f_1(\theta)&=&\fr{c-2}{8} \sin \te \left ( \cos 2 \theta +\fr{1}{3} \right).
\enq
where $c=\fr{\pi^2 N^2 l_s^2}{r^3}$.
It is interesting to notice that $f_1(\theta) \sim \sin \te P_2(\cos \theta) $, and thus it represents a quadrupole distribution of charge.

We now have all the ingredients to find the total potential to second order in $\e$. Expanding, we obtain the new equilibrium radius and energy for small $\e$
\beq
r_{equilibrium}&=&r_{*} \left (1-\fr{\e}{6} \right ), \\
E&=&\frac{2 r_{*} K}{\pi l_s^2} \left (1- \fr{1}{45} \e^2 \right ), \no
\enq
where $r_{*}={ (\pi N)^{2/3} l_s}/{2}$ is the radius of the spherical system. For $\e>0$/ $\e<0$ the system lowers its energy by  shrinking / expanding and squashing/pancaking.  The only contribution  to the energy appears to order $\e^2$ and is negative.  The system suggested in \cite{bbst} as a Quantum Hall soliton is not in static equilibrium classically. 

\subsection{Spherical shift}
%\begin{right}
\begin{figure}[hc]
\scalebox{0.5}{\includegraphics{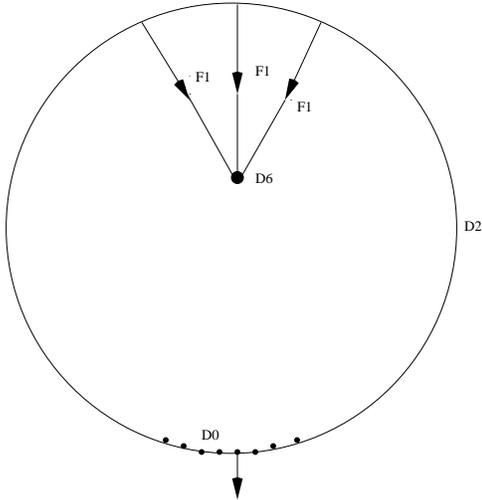}}
\caption{The shifting of the D2 sphere relative to the D6 core may induce another instability in the Quantum Hall Soliton. }
\end{figure}
%\end{right}
Let us consider a deformation in which the center of the QH soliton is displaced a distance $x =\e r $ from the position of the D6 branes. The spacetime embedding is
\beq
x^1=r\sin \theta \cos \vp,\ \ \  
x^2=r\sin \theta \sin \vp, \ \ \
x^3=r(\e+ \cos \theta). 
\enq
The induced metric and 2-form are now
\beq
G_{00}&=&\sqrt{\fr{\rho}{l_s}},\ \ \ G_{\te\te}=-r^2 \sqrt{\fr{l_s}{\rho}},  \nonumber \\
G_{\vp\vp}&=&-\sin^2 \te r^2 \sqrt{\fr{l_s}{\rho}},\ \ \ 
H_{\theta \vp}=\frac{K l_s \sin \theta (1+\e \cos \te)}{2(1+2 \epsilon \cos \theta + \e^2)^{3/2}},
\label{indmetric2}
\enq
where $\rho=r\sqrt{1+2 \epsilon \cos \theta+\e^2}$ is the distance from the D6 branes.
Substituting the metric, dilaton and field strength, we find the DBI potential 
\be
V_{DBI} = {Kr \over 4 \pi l_s^2} \int_0^{\pi} \sqrt{B(\te)+C(\te)f(\te)^2+\sin^2 \te (\p_{\te}A)^2 },
\label{vdbi2}
\ee
where now $B(\te)= {\sin^2 \te \over 1+2 \e \cos \te+\e^2}$, $C(\te)=\fr{\pi^2 N^2 l_s^2}{r^3 (1+2 \e \cos \te+\e^2)^{1/2}}$. The Wess-Zumino potential is given by (\ref{vwz}), where now $H(\te)= \sin \theta (1+\e \cos \te)(1+2 \epsilon \cos \theta + \e^2)^{-3/2}$.

By taking functional derivatives one obtains again equations (\ref{main1},\ref{main2},\ref{main3}). Similarly $V_{strings}$ gives a zero contribution to the potential. Also, only the first order correction to $f(\te)$ contributes to the second order correction to the energy. This correction can be found in a similar fashion to be:
\be
f_1(\theta)= {c \over 2}  \sin \te \cos \te.
\en
This correction represents a dipole perturbation.
The total potential can be easily found. For a fixed $\e$, the energy is minimized by:
\beq
r&=&r_* (1-2 \e^2 ) \nonumber \\
V&=&{2 r_* K \over \pi l_s^2} (1-{\e^2\over 3})
\label{vpot}
\enq
Again there is no $O(\e)$ contribution, and the $\e^2$ contribution is negative. Thus this perturbation is also tachyonic.

If one ignores possible effects of magnetic interactions, one can find a time scale associated with the naive instability, by comparing the kinetic energy of the system to the potential energy. If the D2 brane moves with a velocity $v={\p x \over \p \tau} $, the $G_{00}$ component of the pullback of the metric will receive a correction of the form $ \delta G_{00} \sim G_{xx} \left({\p x \over \p \tau }\right)^2$, which will result in a correction to the energy
\be
K_E = \delta  E \sim V {G_{xx} v^2 \over G_{00}},
\ee
where $V$ is the energy of the static configuration. Evaluating this correction at the position of the D2 brane we obtain
\be
K_E \sim  V N^{-2/3} v^2
\label{KE}
\ee
Comparing this kinetic energy with the potential energy (\ref{vpot}), and ignoring all magnetic effects this gives a naive time scale for the instability $\tau \sim N^{1/3}$. Translated in the proper frame of the D2 brane this gives a time scale 
\be
T_{B=0}=l_s N^{1/2}, \label{T} 
\ee
of the same order as the time scale associated with QH physics.

\section{A dynamical analysis}
In the above chapters, we have found that the potential energy of the QH soliton decreases under displacement. This rules out a static equilibrium configuration. Nevertheless, since the system contains magnetic forces (coming from the interaction of the string ends with D0 branes and from the D0-D6 interactions), it could still be stabilized dynamically  \footnote{We ignored the effects of magnetic interactions in the first version of this paper. We thank L. Susskind for bringing them to our attention.}. The complete analysis of the stabilization is involved and is being presently pursued \cite{lenny}.

One can understand the physics of this dynamical stabilization mechanism by considering a particle of charge $e$ and mass $m$ in an upside down harmonic oscillator ($V=-k x^2$) in a magnetic field $B$ \cite{lenny}. For large enough $B$, the particle does not escape to infinity but rather circles around the origin.  There are two possible orbits in which the repulsive and centrifugal forces are balanced by the magnetic force. The wobbling frequencies are given by the equation
\be
m \w^2 \pm e B \w + k =0, \label{equation}
\ee
whose solutions are
\be
\w_{1,2} ={e B \over 2m } \pm {\sqrt{e^2 B^2 -4 k m} \over 2m}.
\label{solutions}
\ee 
We can see that if 
\be
e^2 B^2 >  4 km \label{condition}
\ee
the solutions are oscillatory (the magnetic field is strong enough to keep the particle from escaping). If (\ref{condition}) is not satisfied, the particle escapes, which corresponds to the collapse of the QH soliton.
In the previous chapter we have found the time scale associated with the decay of the QH soliton in the absence of magnetic field:
\be
\w_{decay} = {1 \over T_{B=0}} \sim {l_s^{-1} N^{-1/2}}.
\label{w}
\ee
In the toy problem of the upside down harmonic oscillator this corresponds to $\w_{decay}= \sqrt{k \over m}$. 

There are three cases to consider. If $e^2 B^2 <  4 km$, the system collapses. If  $e^2 B^2 $ and  $4 km$ are of the same order, the two possible wobbling frequencies will be of the same order as $\w_{decay}$. This frequency is the same as the frequency associated with excited long strings and D2 brane modes. Moreover, it is above the Quantum Hall energy scale \cite{bbst}. Thus the wobbling will dump energy into these modes, and will make the system  unsuitable for describing Quantum Hall physics.

The third case, $e^2 B^2 \gg  4 km$ offers a bit more hope. In the large $B$ limit one of the modes has very high frequency while the other has very low frequency compared to $\w_{decay}$. After the high frequency mode dumps its energy into string modes, the system will be left circling around the origin with a period far larger than the time scale at which Quantum Hall physics occurs.

As we mentioned before, taking into account all magnetic effects is currently under investigation \cite{lenny}. Based on the intuition \cite{bbst} that all time scales in the system are of order $l_s^{-1} N^{-1/2}$, we believe that magnetic effects also involve time scales of the order of $\w_{decay}$, and thus the system falls in the second category. The crude estimate below shows that this may be indeed the case.

There are two sources for the magnetic effect which might stabilize the brane \cite{lenny}. The first is D0 - string end interaction, and the second is D0-D6 interaction. The time scale associated with the first interaction is the cyclotron frequency of the string ends. This was found \cite{bbst} to be
$\w_{cyclotron}= T^{-1}_{cyclotron} = l_s^{-1} N^{-1/2}$. This frequency is of the same order as $\w_{decay}$. The second term comes from a contribution to the Wess-Zumino action of the form
\be
\int C_{0} F_{12},
\ee
which is nonzero when the brane moves. One can easily estimate the contribution of this term to the energy to be
\be
\delta E \sim V {N v x\over r^3}.
\ee
by comparing this energy with (\ref{KE}), we obtain a time scale of the same order as $\w_{decay}$. Therefore, we expect that generic magnetic effects will cause a wobbling with a frequency too large to allow for the modeling of Quantum Hall physics. It is interesting to see if in the exact analysis this will indeed be the case.

\section{A possible stabilization mechanism}

In the stability analysis done above, we studied the system starting from the classical ground state of the spherical configuration, in which string ends do not move on the D2 brane. If string ends were moving they would classically  emit gravitational and antisymmetric tensor radiation, and the system will return to the ground state.

Nevertheless, we can argue that quantum mechanically the string ends move in magnetic field even in their ground state. The motion of a string end can enclose one or more flux quanta. Thus, the string ends and some of the D0 branes do not move independently. 

One can argue further that the fractional charge quasiparticles appearing at filling fractions $\nu \equiv K/N$ when $1/\nu$ is odd, can be thought of as being composites made of an electron and $1/\nu$ magnetic vortices. Thus, in an effective description of the system at these filling fractions the strings and the D0 branes move together. 

As one might imagine, if strings and D0 branes are not moving independently, the intuitive picture of the instability is no longer valid. It is now impossible for strings to move towards one part of the brane and for D0 branes to move towards another part.

One needs to do a computation to find if the configuration is still unstable when one imposes this extra constraint. We only examine what happens to the spherical shift instability in the case when all the D0 branes move with the string ends.   

If we assume that the density of string ends and $F_{\te\vp}$ are both proportional to $f(\te)$, the potential which one has to minimize becomes 
\beq
V_{frozen} = {K r \over 4 \pi l_s^2} \left[ \int{ d \te \ {\rho \over r} f(\te) + A(\te) f(\te) - H(\te) A(\te)}  \right. \nonumber \\
\left. + \int{d \te \sqrt{B(\te)+\sin^2 \te \left(\p_{\te}A(\te)\right)^2 + C(\te) f(\te)^2}}\  \right]
,
\label{vfrozen}
\enq
where the functions $\rho,B(\te),C(\te)$, and $H(\te)$ are given in Section 3.3. The function $f(\te)$ satisfies the constraint $\int f(\te) = 2$. By taking functional derivatives of (\ref{vfrozen}) we obtain two equations relating $f(\te)$ and $A(\te)$. 
\beq
f(\te)-H(\te) &=& \p_{\te} {\sin^2 \te \p_{\te} A(\te)\over \sqrt{B(\te)+\sin^2 \te \left(\p_{\te}A(\te)\right)^2 + C(\te) f(\te)^2}},  \label{m1} \\
\lambda + {\rho \over r} + A(\te) &=& {- C(\te) f(\te) \over \sqrt{B(\te)+\sin^2 \te \left(\p_{\te}A(\te)\right)^2 + C(\te) f(\te)^2}}, \label{m2}
\enq
where $\lambda$ enforces the constraint and $A(\te)$ is defined up to a constant. As before, to find the $\e^2$ contribution to the energy, one only needs to find the first order in $\e$ correction to $f(\te)$ and $A(\te)$, which we call $\e\sin \te \ g(\te)$ and $\e A_1(\te)$. Expanding (\ref{m1},\ref{m2}) and substituting the value of $c$ at $r_*$, we obtain two simpler equations:
\beq
3(g(\te)+2 \cos \te)&=&{1 \over \sin \te}\p_{\te}(\sin \te \p_{\te} A_1),  \nonumber \\
27 A_1(\te)&=& 5 \cos \te -8 g(\te),
\label{ecfinal}
\enq
which can be solved to give
\be
A(\te)=\e {63 \over 65} \cos \te, \ \ \ \ f(\te) = \sin \te (1 - \e {172 \over 65} \cos \te),
\ee
where we discarded the homogeneous solutions coming from (\ref{ecfinal}) for being singular, and thus unphysical.
The total energy of the system at $r=r_{*}$ is:
\beq
V_{frozen}={K r_{*} \over 2 \pi l_s^2}[4 +{374 \over 585}\e^2], \nonumber \\
r=r_{*}(1+.78 \e^2).
\enq 
We observe that the correction to $f(\te)$ has a dipole form. Nevertheless, it has a sign opposite to that of the correction obtained for $f(\te)$ in Section $3.3\ .$ 

We have found that under the assumption that string ends and D0 branes move together the system is stable under a dipole perturbation. Based on the intuition that dipole corrections to the energy  are larger than those of higher moments, this seems to indicate that the system is always stable when the string ends and D0 branes move together.  

\section{Conclusions}

We have computed the energy shift of the Quantum Hall Soliton upon displacement, and found it negative. Classically this rules out a static equilibrium configuration. Nevertheless, since the system contains magnetic fields, this does not rule out dynamical stability. A rough estimate of the strength of magnetic effects indicates that even if a dynamical stabilization mechanism is possible, it probably makes the system unsuitable for modeling Quantum Hall physics.

We have also proposed a stabilization mechanism based on the intuition that in a quantum treatment of the problem, string ends have to circle around one or more magnetic flux quanta. We have found that under the assumption that string ends and the D0 branes move together, the system is stable.

\section{Acknowledgements}
We would like to thank Joe Polchinski, Leonard Susskind, Andrei Mikhailov, Leon Balents and Smitha Vishveshwara for discussions.
This work was supported by NSF grant PHY97-22022 and DOE contract DE-FG-03-91ER40618.


\begin{thebibliography}{99}
\bibitem{bbst} B.A. Bernevig, J. Brodie, L. Susskind and N. Toumbas hep-th/0010105.
\bibitem{hw} A. Hanany and E. Witten, Nucl. Phys. B.492 (1997) 152, hep-th/9611230
\bibitem{burgess} C.P. Burgess, Nucl. Phys. B294 (1987) 427
\bibitem{lenny} L. Susskind - private communication
\end{thebibliography}
\end{document}